\documentstyle[preprint,prd,aps]{revtex}
\newcommand{\be}{\begin{equation}}
\newcommand{\ee}{\end{equation}}
\newcommand{\bea}{\begin{eqnarray}}
\newcommand{\eea}{\end{eqnarray}}
\begin{document}
\draft
\title{\large\bf  The Akulov-Volkov Lagrangian, Symmetry Currents  \\
and Spontaneously Broken Extended Supersymmetry}
\author{T.E. Clark\footnote{e-mail address: clark@physics.purdue.edu} and S.T. Love\footnote{e-mail address: love@physics.purdue.edu}}
\address{\it Department of Physics, 
Purdue University,
West Lafayette, IN 47907-1396}
\def\overlay#1#2{\setbox0=\hbox{#1}\setbox1=\hobx to \wd0{\hss #2\hss}#1%
\hskip -2\wd0\copy1}
\maketitle
\begin{abstract}
A generalization of the Akulov-Volkov effective Lagrangian governing the self interactions of the Nambu-Goldstone fermions associated with spontaneously broken extended 
supersymmetry as well as their coupling to matter is presented and scrutinized.  The resulting currents associated with R-symmetry, supersymmetry
and space-time translations are constructed and seen to form a supermultiplet structure.  \end{abstract}
\vskip2pc
\pagebreak

\section{Introduction}

Many proposed fundamental theories of nature require there to be supersymmetry (SUSY) in extra dimensions.  When reduced to four dimensions, this results in models exhibiting an extended supersymmetry, $N \geq 2$. Since the low energy standard model of elementary particle interactions has no supersymmetry, $N=0$, the
question of the nature of the chain of supersymmetry beakdown from higher $N$ to $N=0$ is quite relevant. In addition, for those models with $N=1$ supersymmetry at the electroweak scale, the breakdown of SUSY from a higher $N>1$ to $N=1$, either in stages (partial supersymmetry breaking), or directly, is of paramount importance.  Models of partial supersymmetry breaking have been extensively studied and require exploiting one of two possible avenues of realization. Either the supersymmetry current algebra requires the inclusion of certain central charges, \cite{pol}-\cite{taylor}, or when embedded in a supergravity model, the gravitino gauge field contains negative norm states \cite{gira}-\cite{ferr}.  In this paper, we present the construction of the Akulov-Volkov effective action \cite{AV} describing the spontaneous breakdown of $N$-extended supersymmetry to $N=0$ \cite{West}\cite{Nishino}. This model contains only the Nambu-Goldstone fermions necessary for it to describe the spontaneous breakdown of the extended supersymmetry.  Since the supersymmetry charges are in the fundamental representation of an accompanying (unbroken) $SU(N)_R$ symmetry, their Akulov-Volkov realization depends on a single scale, the common Goldstino decay constant.  This fact is explicitly exhibited by exploiting this non-Abelian $SU(N)_R$ symmetry of the $N$-extended SUSY algebra which manifests itself through a rescaling invariance of the action resulting in its single scale dependence. The Noether currents associated with the chiral $SU(N)_R$ symmetries, the $N$ supersymmetries and the space-time translation symmetries are constructed and the supersymmetry transformations of the currents are determined.  In particular the $R$-currents, supersymmetry currents and the energy-momentum tensor are shown to form the components of a supercurrent \cite{FZ}-\cite{CPS}. This extends previous work \cite{CL1} for $N=1$ supersymmetry.  Finally, the couplings of the Goldstino fields to matter and gauge fields are delineated once again generalizing the construction of the $N=1$ case \cite{CL2}-\cite{CL3}.

The $N$-extended supersymmetry Weyl spinor charges $Q_\alpha^A$ and $\bar Q^A_{\dot\alpha}$, where $A=1,2,\ldots , N$, obey the supersymmetry algebra (in the absence of any central charges)
\bea
\left\{Q^A_{\alpha },  \bar Q_{B\dot\alpha}\right\} &=& 2 \delta_B^A \sigma^\mu_{\alpha\dot\alpha} P_\mu \cr
\left\{Q^A_{\alpha},  Q_{\beta}^B\right\} &= 0 =& \left\{\bar Q_{A\dot\alpha},  \bar Q_{B\dot\beta}\right\} \cr
\left[ P^\mu , Q_{\alpha}^{A} \right] &= 0 =& \left[ P^\mu , \bar Q_{A\dot\alpha}\right]
\label{SA}
\eea
with $P^\mu$ the energy-momentum operator.   

This algebra can be realized by means of the Akulov-Volkov nonlinear transformations \cite{AV} of spontaneously broken $N$-extended supersymmetry.  The $N$ Weyl spinor Goldstino fields $\lambda^{\prime\alpha}_A (x)$ and their hermitean conjugate fields $\bar\lambda^{\prime A}_{\dot\alpha} (x)$ are defined to transform under the nonlinear extended supersymmetry as
\bea
\delta^Q (\xi,\bar\xi) \lambda^{\prime \alpha}_A  &=& f_A\xi^\alpha_A +\Lambda^{\rho} (\xi, \bar\xi)\partial_\rho \lambda^{\prime \alpha}_A \cr
\delta^Q (\xi,\bar\xi) \bar\lambda^{\prime A}_{\dot\alpha}  &=& f_A\bar\xi^A_{\dot\alpha} +\Lambda^{\rho} (\xi, \bar\xi)\partial_\rho \bar\lambda^{\prime A}_{\dot\alpha},
\label{NSUSYP}
\eea
where $\xi^\alpha_A$, $\bar\xi^A_{\dot\alpha}$ are Weyl spinor $N$-SUSY transformation parameters, $f_A$ are (for the moment independent) constants with dimension ${\rm mass}^2$  and 
\be
\Lambda^{\rho} (\xi, \bar\xi) = \frac{1}{if_A}\left[ \lambda^\prime_A \sigma^\rho \bar\xi_A -\xi_A \sigma^\rho \bar\lambda^\prime_A \right] .
\ee
[Comment on notation: Throughout this paper, a summation convention is employed in which all repeated indices in a single term are summed over. The only exceptions are the first terms on the right hand side of Eq. (\ref{NSUSYP}) and Eq. (\ref{RA}). In addition, we shall often supress the indices when they are summed over. Thus, for example, $\xi \sigma^\mu \bar{\xi} \equiv \xi^\alpha_A \sigma^\mu_{\alpha\dot{\alpha}}\bar{\xi}^{A\dot{\alpha}}$.]\\
Further representing the space-time translations as 
\bea
\delta^P (a) \lambda^{\prime \alpha}_A &=& a^\rho \partial_\rho \lambda^{\prime \alpha}_A\cr
\delta^P (a) \bar\lambda^{\prime A}_{\dot\alpha} &=& a^\rho \partial_\rho \bar\lambda^{\prime A}_{\dot\alpha},
\label{DP}
\eea
it is readily established that the above extended SUSY algebra is indeed satisfied by the associated variations.

Using these Goldstino field extended SUSY transformations, it is straightforward to generalize the Akulov-Volkov $N=1$ SUSY construction to form the extended SUSY invariant action, $\Gamma_{AV}=\int d^4x {\cal L}_{AV}$, where 
\be
{\cal L}_{AV}=-\frac{f^2}{2} \det A
\ee
where $f$ is a scale with dimension ${\rm mass}^2$. The Akulov-Volkov vierbein $A_\mu^{~\nu}$ is the $4\times 4$ matrix defined as
\be
A_\mu^{~\nu} = \delta_\mu^{~\nu} + \frac{i}{f_A^2}\left( \lambda^\prime_A \stackrel{\leftrightarrow}{\partial_\mu} \sigma^\nu  \bar\lambda^\prime_A \right).
\ee
Expanding the determinant, a canonically normalized kinetic term for each of the Goldstino fields is secured after the rescalings 
\be
\lambda^\prime_A =\frac{f_A}{f}\lambda_A .
\label{RA}
\ee 
Since all self interactions of the Goldstinos have the functional form (in a short hand notation) $f^2 F(\frac{\lambda^\prime_A \partial \bar{\lambda}^\prime_A}{f_A^2}, \frac{\partial {\lambda}^\prime_A \bar{\lambda}^\prime_A}{f_A^2})= f^2 F(\frac{\lambda_A \partial \bar{\lambda}_A}{f^2},\frac{\partial {\lambda}_A \bar{\lambda}_A}{f^2})$, it follows that there is really only one scale in the action and the Akulov-Volkov Lagrangian can be written as 
\be
{\cal L}_{AV}=-\frac{f^2}{2} \det A
\label{AVL}
\ee
where 
\be
A_\mu^{~\nu} = \delta_\mu^{~\nu} + \frac{i}{f^2}\left( \lambda_A\stackrel{\leftrightarrow}{\partial_\mu} \sigma^\nu  \bar\lambda_A \right).
\ee
Since this Akulov-Volkov action is the unique extended SUSY invariant structure containing the lowest mass dimension Goldstino self interactions, it follows that the extended SUSY algebra is completely broken at this single scale, $\sqrt{f}$, which can be identified as the common Goldstino decay constant \cite{West}. That is, one cannot sequentially break the various supersymmetries at different scales and still realize the algebra of Eq. (\ref{SA}).  This conclusion is in accord with a heuristic argument \cite{W} which follows from the form of the extended SUSY algebra, Eq. (\ref{SA}). If one demands that the Hilbert space of states be positive definite, then the algebra dictates that it cannot be that some of the supersymmetry charges annihilate the vacuum while others do not. As was previously pointed out \cite{pol}, however, this argument is purely formal since for spontaneously broken symmetries, the associated symmetry charges really do not exist. In our approach, we reach an identical conclusion using very concrete effective Lagrangian techniques which explicitly encapsulate the relevant dynamics. The result is certainly a very strong constraint and is very different from the situation often encountered involving multiple global symmetries which can generally  be broken at different scales. Thus we encounter yet another example \cite{CLDIL} of how supersymmetry very tightly constrains the allowed dynamics of a model.

Further note that in terms of the rescaled (unprimed) fields, the extended nonlinear SUSY transformations take the form
\bea
\delta^Q (\xi,\bar\xi) \lambda^\alpha_A  &=& f\xi^\alpha_A +\Lambda^\rho (\xi, \bar\xi)\partial_\rho \lambda^\alpha_A \cr
\delta^Q (\xi,\bar\xi) \bar\lambda^A_{\dot\alpha}  &=& f\bar\xi^A_{\dot\alpha} +\Lambda^\rho (\xi, \bar\xi)\partial_\rho \bar\lambda^A_{\dot\alpha},
\label{NSUSY}
\eea
with
\be
\Lambda^\rho (\xi, \bar\xi) = \frac{1}{if}\left[ \lambda_A \sigma^\rho \bar\xi_A -\xi_A \sigma \bar\lambda_A \right] .
\ee
Clearly, the supersymmetry algebra
\bea
\left[\delta^Q (\xi, \bar\xi), \delta^Q (\eta, \bar\eta)\right] \lambda^\alpha_A &=& -2i \delta^P (\xi_B\sigma\bar\eta^B -\eta_B\sigma\bar\xi^B) \lambda^\alpha_A \cr
\left[\delta^Q (\xi, \bar\xi), \delta^Q (\eta, \bar\eta)\right] \bar\lambda^A_{\dot\alpha} &=& -2i \delta^P (\xi_B\sigma\bar\eta^B -\eta_B\sigma\bar\xi^B) \bar\lambda^A_{\dot\alpha} \eea
continues to be satisfied. Here the space-time variations, $\delta^P(a)$, of the unprimed fields are of the same form as for the primed fields (cf. Eq. (\ref{DP})).

The Akulov-Volkov Lagrangian is invariant while the Goldstino SUSY transformations are covariant under a global $SU(N)$ symmetry which has the properties of an $R$ symmetry. Under this $SU(N)_R$ symmetry, the Goldstino fields transform as
\bea
\delta^R (\omega) \lambda^\alpha_{A} &=& -i\lambda^\alpha_{B} \omega_a\left( {T}^a \right)^B_{~A} \cr
\delta^R (\omega) \bar\lambda^A_{\dot\alpha} &=& i \omega_a\left( {T}^a \right)^A_{~B}  \bar\lambda^B_{\dot\alpha} ,
\label{Rtrans}
\eea
where $\omega_{a}$, $a = 1, 2, \ldots , N^2 -1$, are the $SU(N)_R$ transformation parameters and $T^a$ denote the fundamental representation matrices of $SU(N)$. It follows that 
\bea
\left[ \delta^R (\omega), \delta^Q (\xi, \bar\xi) \right] \lambda^\alpha_A  &=& \delta^Q (\xi_R, \bar\xi_R) \lambda^\alpha_A \cr
\left[ \delta^R (\omega), \delta^Q (\xi, \bar\xi) \right] \bar\lambda^A_{\dot\alpha}  &=& \delta^Q (\xi_R, \bar\xi_R) \bar\lambda^A_{\dot\alpha} , 
\eea
which is precisely what is required of an $R$-symmetry. Here 
\bea
\xi^\alpha_{RA} &\equiv& i \xi^\alpha_B \omega_a\left( {T}^a \right)^B_{~A} \cr
\bar{\xi}_{R\dot\alpha}^{A} &\equiv& -i \omega_a\left( {T}^a \right)^A_{~B} \bar\xi^B_{\dot\alpha} .
\label{XiR}
\eea

This symmetry is also reflected in the extended SUSY algebra, Eq. (\ref{SA}), which remains invariant under an $SU(N)$ rotation of the supersymmetry charges. The $Q^A$ and $\bar Q_A$ transform as the $N$ and $\bar N$ fundamental representations of this $SU(N)_R$ symmetry,  so that  
\bea
\left[ R^a, Q^B \right] &=& i \left( T^a \right)^B_{~C} Q^C \cr
\left[ R^a, \bar Q_B \right] &=& -i\bar Q_C \left( T^a \right)^C_{~B} ,
\label{Ralg}
\eea
while the energy-momentum operator is $R$ invariant:
\be
[R^a , P^\mu ]=0 .
\ee
It should be noted that Eq. (\ref{Ralg}) requires the Akulov-Volkov realization of the SUSY transformations to form fundamental representations of $SU(N)_R$ which, along with Eq. (\ref{Rtrans}), implies that the Goldstino decay constants must all be the same.

Combining these various results, it follows that the Akulov-Volkov Lagrangian \cite{West}\cite{Nishino} of Eq. (\ref{AVL}) is invariant under $SU(N)_R$ transformations while transforming as a total divergence under SUSY and space-time translations:
\bea
\delta^R (\omega) {\cal L}_{AV} &=& 0 \cr
\delta^Q (\xi, \bar\xi) {\cal L}_{AV} &=& \partial_\rho \left[ \Lambda^\rho (\xi, \bar\xi) {\cal L}_{AV}\right] \cr
\delta^P (a) {\cal L}_{AV} &=& \partial_\rho \left[ a^\rho {\cal L}_{AV}\right]  .
\eea
Hence the action constructed from this Lagrangian is invariant under these transformations.

\section{Symmetry Currents}

Noether's theorem can be used to construct the conserved currents associated with the above symmetries of the action. First introduce the local variation by means of the functional differential operator
\be
\delta (x) = \zeta^\alpha_A (x) \frac{\delta}{\delta \lambda^\alpha_A (x)} + \bar\zeta^{A \dot\alpha } (x) \frac{\delta}{\delta \bar\lambda^{A\dot\alpha} (x)} ,
\ee
which allows for space-time and field dependent Weyl spinor variation parameters, $\zeta (x)$ and $\bar\zeta (x)$. When applied to the Akulov-Volkov Lagrangian, Eq. (\ref{AVL}), this yields
\bea
\delta(y) {\cal L}_{AV}(x) &=& \frac{f}{2}\det{A(x)} (A^{-1})_\nu^{~\mu} (x)\partial_\mu^x \left[ \delta^4 (x-y)\Lambda^\nu (\zeta(x) ,\bar\zeta(x) ) \right] \cr
 & & - i\delta^4 (x-y) \left[\det{A(x)}(A^{-1})_\nu^{~\mu} (x)\left(\zeta_A(x) \sigma^\nu \partial_\mu \bar\lambda^A(x) -\partial_\mu \lambda_A(x)  \sigma^\nu \bar\zeta^A(x) \right) \right] .
\label{LVL}
\eea
Further defining the variation operator, 
\be
\delta = \int d^4 x \delta (x) ~,
\ee
it follows that 
\be
\delta {\cal L}_{AV}(x) = -\frac{i}{2} \det{A}(x) (A^{-1})_\nu^{~\mu} (x)\left[ \zeta_A (x)\stackrel{\leftrightarrow}{\partial_\mu}  \sigma^\nu \bar\lambda^A(x) +\lambda_A (x)  
\stackrel{\leftrightarrow}{\partial_\mu} \sigma^\nu \bar\zeta^A (x) \right] ,
\label{VL}
\ee
while the local variation of the Akulov-Volkov action, $\Gamma_{AV}=\int d^4x {\cal L}_{AV}$, is secured as  
\be
\delta (x) \Gamma_{AV} = \delta {\cal L}_{AV}(x) - \partial_\mu \left[ \frac{f}{2}\det{A(x)} 
(A^{-1})_\nu~^\mu (x)\Lambda^\nu (\zeta (x), \bar\zeta (x)) \right] 
\label{Noether}
\ee
which constitutes Noether's theorem.

To extract the various currents and their conservation laws, one allows the local transformation parameters $\zeta^\alpha_A$ and $\bar\zeta^A_{\dot\alpha}$ to assume the various forms of the associated Goldstino transformation laws. For example, taking the explicit form to be that of a space-time translation of the Goldstino field
\bea
\zeta^\alpha_A (x)&=& a^\rho \partial_\rho \lambda^\alpha_A (x)\cr
\bar\zeta^A_{\dot\alpha} (x)&=& a^\rho \partial_\rho \bar\lambda^A_{\dot\alpha} (x)
\label{translation}
\eea
and substituting into Noether's theorem, Eq. (\ref{Noether}), leads to the conserved Noether energy-momentum tensor 
\be
T^\mu_{~\nu}(x) = -\frac{f^2}{2} (\det{A}(x)) (A^{-1})_{\nu}^{~\mu}(x).
\ee
and its conservation law
\be
a_\nu\partial_\mu T^{\mu\nu}(x) = \delta^{P} (x;a) \Gamma_{AV} .
\ee
with
\be
\delta^{P} (x;a) = a_\mu[\partial^\mu \lambda^\alpha_A (x) \frac{\delta}{\delta \lambda^\alpha_A (x)} + \partial^\mu \bar\lambda^{A \dot\alpha } (x) \frac{\delta}{\delta \bar\lambda^{A\dot\alpha} (x)}] .
\ee

Similarly, the parameters can be chosen as 
\bea
\zeta^\alpha_A (x)&=& -i \lambda^{B\alpha} (x)\omega_a \left(  {T}^a \right)^B_{~A} \cr
\bar\zeta^{A \dot\alpha} (x)&=& i \omega_a\left( {T}^a \right)^A_{~B} \bar\lambda^{B \dot\alpha} (x) 
\eea
which correspond to the form of a Goldstino $SU(N)_R$ transformation. This time substitution into Noether's theorem produces the conserved $R$ current 
\be
R^\mu (\omega) \equiv \omega_{a} R^{{a}\mu}(x) = \frac{2}{f^2} T^\mu_{~\nu}(x) \left[ \lambda_A(x) \omega_a \left({T}^a \right)^A_{~B} \sigma^\nu \bar\lambda^B (x)\right] .
\ee
and its conservation law
\be
\partial_\mu R^{\mu}(\omega)  = \delta^R (x;\omega) \Gamma_{AV} ,
\ee
where the Ward identity functional differential operator describing $R$ transformations is  given by
\be
\delta^R (x;\omega) = \omega_a\left[-i(T^{a})^C_{~B}\lambda_C^{\alpha}(x) \frac{\delta}{\delta \lambda^\alpha_B (x)} +
i(T^{a})^B_{~C} \bar\lambda^{C \dot\alpha}(x) \frac{\delta}{\delta \bar\lambda^{B\dot\alpha} (x)}\right].
\ee

Finally, the parameters can be chosen to have the form of the Goldstino nonlinear SUSY transformations 
\bea
\zeta_A^\alpha (x) &=& f \xi^\alpha_A + \Lambda^\rho (\xi, \bar\xi) \partial_\rho \lambda^\alpha_A (x)\cr
\bar\zeta^{A \dot\alpha} (x) &=& f \bar\xi^{ A\dot\alpha} + \Lambda^\rho (\xi, \bar\xi) \partial_\rho \bar\lambda^{ A\dot\alpha} (x),
\eea
with the $\xi$ and $\bar\xi$ appearing on the right hand side being the usual constant spinor SUSY transformation parameters.  So doing, the supersymmetry currents $Q^\mu_{A\alpha}$ and $\bar Q^\mu_{A\dot\alpha}$  are obtained from Noether's theorem as
\be
Q^\mu (\xi, \bar\xi) = 2T^\mu_{~\nu}(x)\Lambda^\nu (\xi, \bar\xi) ,
\ee
where $Q^\mu (\xi, \bar\xi) = \xi^\alpha_A Q^\mu_{A\alpha}(x) + \bar Q^\mu_{A\dot\alpha}(x) \bar\xi^{\dot\alpha A}$.  The current conservation
equation takes the form
\be
\partial_\mu Q^\mu (\xi, \bar\xi) = \delta^Q (x;\xi,\bar\xi) \Gamma_{AV} ,
\ee
with
\bea
\delta^Q (x;\xi, \bar\xi) &=& \left(f \xi^\alpha_A + \Lambda^\rho (\xi, \bar\xi) \partial_\rho \lambda^\alpha_A (x)\right)  \frac{\delta}{\delta \lambda^\alpha_A (x)} \cr
 & & + \left(f \bar\xi^{A \dot\alpha} + \Lambda^\rho (\xi, \bar\xi) \partial_\rho \bar\lambda^{A \dot\alpha} (x) \right) \frac{\delta}{\delta \bar\lambda^{A\dot\alpha} (x)}.
\eea

The conserved currents in the effective theory are related through their SUSY transformation properties.  The SUSY transformation, $\delta^Q(\xi,\bar\xi)=\int d^4x \delta^Q(x;\xi,\bar\xi)$, of the $R$ current produces the supersymmetry current plus a term that is a Belinfante improvement term for the supersymmetry currents as well as additional Euler-Lagrange equation terms:
\be
\delta^Q (\xi, \bar\xi) R^\mu (\omega) = -Q^\mu (\xi_R, \bar\xi_R) +\partial_\rho q^{\rho\mu} (\omega, \xi, \bar\xi) +\Lambda^\mu (\xi, \bar\xi) \partial_\rho R^\rho (\omega).
\ee
Here the $\rho\mu$ antisymmetric improvement terms are defined as
\be
q^{\rho\mu} (\omega, \xi, \bar\xi) = \Lambda^\rho (\xi, \bar\xi) R^\mu (\omega) - \Lambda^\mu (\xi, \bar\xi) R^\rho (\omega) .
\ee
Note that on-shell, the divergence of the $R$ current vanishes as a consequence of the $R$-current conservation law and the field equations. 

Analogously, the Noether SUSY currents transform into the Noether energy-momentum tensor, its Belinfante improvements, other trivially conserved (without need of field equations) antisymmetric terms and Euler-Lagrange equation terms that enter through the divergence of the supersymmetry currents which again vanish on-shell:
\be
\delta^Q (\eta, \bar\eta) Q^\mu (\xi, \bar\xi) = 2i \left( \xi \sigma_\nu \bar\eta - \eta \sigma_\nu \bar\xi \right) T^{\mu\nu} + \partial_\rho G^{\rho\mu} (\eta, \bar\eta, \xi, \bar\xi ) +\Lambda^\mu (\eta, \bar\eta) \partial_\rho Q^\rho (\xi, \bar\xi) .
\ee
The $\rho\mu$ antisymmetric improvement terms $G^{\rho\mu}$ are defined as
\be
G^{\rho\mu} (\eta, \bar\eta, \xi, \bar\xi )  = \Lambda^\rho (\eta, \bar\eta) Q^\mu (\xi, \bar\xi) - \Lambda^\mu (\eta, \bar\eta) Q^\rho (\xi, \bar\xi ).
\ee

Finally the SUSY variation of the energy-momentum tensor does not lead to another conserved current but only to a trivially conserved antisymmetric term and the on-shell vanishing Euler-Lagrange equation terms so that
\be
\delta^Q (\xi, \bar\xi) T^{\mu\nu} = \partial_\rho \left[ \Lambda^\rho (\xi, \bar\xi) T^{\mu\nu} -\Lambda^\mu (\xi, \bar\xi) T^{\rho\nu} \right]  + \Lambda^\mu (\xi, \bar\xi ) \partial_\rho T^{\rho\nu} .
\ee
Hence, the currents $R^\mu$, $Q^\mu$, and $T^{\mu\nu}$ form the component currents of  a supercurrent.  Given the $R$ current, the SUSY currents and the energy-momentum tensor can be defined by SUSY variations of $R^\mu (\omega)$ through the above transformation equations.  Due do the $R$ invariance of the $N$ extended SUSY algebra, all of the lower component $R$ and SUSY currents transform into the energy-momentum tensor. The form of the currents, their conservation laws and multiplet structure exhibit an entirely similar structure to that found for the $N=1$ nonlinear SUSY theory \cite{CL1}.

\section{Matter and Gauge Fields}

The $N$ extended supersymmetry algebra can also be nonlinearly realized on matter (non-Goldstino, non-gauge) fields, generically denoted by $\phi^i$, where $i$ can represent any Lorentz or internal symmetry labels, using the standard realization \cite{CL2},\cite{CL3},\cite{WB}
\be
\delta^Q (\xi, \bar\xi) \phi^i = \Lambda^\rho (\xi, \bar\xi) \partial_\rho \phi^i .
\ee
The matter fields also carry a representation of the chiral $SU(N)_R$ symmetry so that
\be
\delta^R(\omega) \phi^i = i \omega_a \left( {R^a} \right)^i_{~j} \phi^j ,
\ee
with $(R^{a})^i_{~j}$ constituting a set of $SU(N)$ representation matrices. It is straightforward to check that the extended SUSY algebra is obeyed by these transformations:
\bea
\left[ \delta^Q (\xi, \bar\xi), \delta^Q (\eta, \bar\eta) \right] &=& -2i \delta^P (\xi \sigma \bar\eta -\eta \sigma \bar\xi ) \cr
\left[ \delta^Q (\xi, \bar\xi), \delta^P (a) \right] &=& 0 = \left[ \delta^R (\omega), \delta^P (a) \right] \cr
\left[\delta^R (\omega),  \delta^Q (\xi, \bar\xi) \right] &=& -\delta^Q (\xi_R, \bar\xi_R) \cr
\left[\delta^R (\omega),  \delta^R (\theta) \right] &=& \delta^R (\vec\omega \times \vec\theta) ,
\eea
where $\xi_R$ and $\bar\xi_R$ are defined in Eq. (\ref{XiR}) and the  cross product is given in terms of the $SU(N)$ structure constants, $f_a^{~bc}$, so that $(\vec\omega \times \vec\theta )_{a} = f_a^{~bc}\omega_{b} \theta_{c}$.

The formalism required to construct extended SUSY invariant actions involving matter fields, gauge fields and their interactions with Goldstinos is very similar to that previously deduced for the case of $N=1$ supersymmetry. As such, we shall only outline the necessary steps and refer the reader to the literature \cite{CL2},\cite{CL3} to fill in the details.

The SUSY covariant derivative of the matter field, ${\cal D}_\mu \phi$, is defined as
\be
{\cal D}_\mu \phi = (A^{-1})_\mu^{~\nu} \partial_\nu \phi ,
\ee
where $(A^{-1})_\mu^{~\nu}$ is the inverse Akulov-Volkov vierbein, $A_\mu~^\rho 
(A^{-1})_\rho^{~\nu} = \delta_\mu~^\nu$. So doing, it transforms under SUSY as the standard realization so that
\be
\delta^Q (\xi, \bar\xi) ({\cal D}_\mu \phi) = \Lambda^\rho (\xi, \bar\xi) \partial_\rho ({\cal D}_\mu \phi).
\ee

If the matter field belongs to a representation of an internal local symmetry group ${\cal G}$
\be
\delta^G (\theta) \phi^i = i \theta_I( L^I )^i_{~j} \phi^j ,
\ee
where $\theta_I = \theta_I (x)$ are the space-time dependent gauge transformation parameters and $(L^I)^i_{~j}$ are a set of ${\cal G}$ representation matrices, then the gauge covariant derivative is given by
\be
\left( D_\mu \phi \right)^i = \partial_\mu \phi^i +(L^I)^i_{~j} A_{I\mu} \phi^j .
\ee
Here $A_{I\mu}$, $I = 1,2, \ldots, {\rm dim}~{\cal G}$, are the ${\cal G}$ gauge fields having gauge transformation properties
\be
\delta^G (\theta) A_{I\mu} = \left( D_\mu \theta \right)_I = \partial_\mu \theta_I +g f_I^{~JK} A_{J\mu} \theta_K .  
\ee
with $g$ the gauge coupling constant and $f_I^{~JK}$ are the group ${\cal G}$ structure constants. These $A_{I\mu}$ carry the non-standard SUSY realization
\be
\delta^Q (\xi, \bar\xi) A_{I\mu} = \Lambda^\rho (\xi, \bar\xi ) \partial_\rho A_{I\mu} + \partial_\mu \Lambda^\rho (\xi, \bar\xi) A_{I\rho} .
\ee
By once again using the inverse Akulov-Volkov vierbein, 
the gauge covariant derivative of $\phi$ can also be made SUSY covariant via 
\be
\left( {\cal D}_\mu \phi \right)^i = (A^{-1})_\mu^{~\nu} \left( D_\nu \phi \right)^i  ,
\ee
so that it transforms as the standard realization under the extended nonlinear SUSY, $\delta^Q (\xi, \bar\xi) \left( {\cal D}_\mu \phi \right)^i = \Lambda^\rho (\xi, \bar\xi) \partial_\rho \left( {\cal D}_\mu \phi \right)^i $.  The gauge field is chosen to be a singlet under $R$ transformations, $\delta^R (\omega) A_{I\mu} =0$, so the gauge and $R$ transformations commute.  

Alternatively, a redefined gauge field can be introduced as
\be
V_{I\mu} \equiv (A^{-1})_\mu^{~\nu} A_{I \nu}~~,
\ee
so that it transforms as the standard realization
\be
\delta^Q(\xi,\bar{\xi})V_{I \mu}  = \Lambda^\rho (\xi, \bar\xi) \partial_\rho V_{I\mu}
\ee
and the gauge covariant derivative takes the form
\be\label{coderiv}
({\cal D}_\mu \phi)^i \equiv (A^{-1})_\mu^{~\nu} \partial_\nu \phi^i + (L^I)^i_{~j}{V}_{I\mu} \phi^j~~.
\ee
Moreover, the redefined gauge field $V_\mu^I$ transforms under gauge transformations as
\be
\delta^G(\theta) V_{I \mu}  = (A^{-1})_\mu^{~\nu} (D_\nu \theta)_I~~.
\ee
For all realizations, the gauge transformation and SUSY transformation commutator yields a gauge 
variation with a SUSY transformed value of the gauge transformation parameter:
\be
\left[\delta^G(\theta),\delta^Q(\xi, \bar\xi)\right]=\delta^G(\Lambda^\rho (\xi, \bar\xi) \partial_\rho \theta-\delta^Q (\xi, \bar\xi) \theta)~~.
\ee
Alternately, by requiring the local gauge transformation parameter to also transform under the standard realization,
\be
\delta^Q(\xi,\bar{\xi}) \theta_I  = \Lambda^\rho (\xi, \bar\xi) \partial_\rho \theta_I~~,
\ee
then the gauge and SUSY transformations commute.

To construct an invariant kinetic energy term for the gauge fields it is convenient for the anti-symmetric tensor field strength to be brought into the standard realization. This is achieved by defining
\be
\label{field}
{\cal F}_{I\mu\nu} = (A^{-1})_\mu^{~\alpha} (A^{-1})_\nu^{~\beta} F_{I\alpha\beta}~~,
\ee
where  $F_{I\alpha\beta} $ the usual field strength
\be
F_{I\alpha\beta} = \partial_\alpha A_{I\beta} -\partial_\beta A_{I\alpha} +if_I^{~JK}A_{J\alpha} 
A_{K\beta}~.
\ee
Under SUSY transformations, $F_{I\mu\nu}$ varies as 
\be
\delta^Q(\xi,\bar{\xi}) F_{I\mu\nu} = \Lambda^\rho\partial_\rho F_{I\mu\nu}  +\partial_\mu 
\Lambda^\rho F_{I\rho\nu} +\partial_\nu \Lambda^\rho F_{I\mu\rho} ~~.
\ee
while
\be
\delta^Q(\xi,\bar{\xi}) {\cal F}_{I\mu\nu} = \Lambda^\rho\partial_\rho {\cal F}_{I\mu\nu} ~~.
\ee
These standard realization building blocks, the Akulov-Volkov vierbeine, $A_\mu~^\nu$, $(A^{-1})_\mu^{~\nu}$, the covariant derivatives, ${\cal D}_\mu \phi^i,~ {\cal D}_\mu \lambda_A ,~ {\cal D}_\mu \bar\lambda^A$ and the field strength tensor, ${\cal F}_{I\mu\nu}$, and higher covariant derivatives thereof, can be  combined to make SUSY and gauge invariant actions.

\section{Invariant Actions}

SUSY and gauge invariant actions can be constructed using the fact that the matter fields and their covariant derivatives transform according to the standard realization. The final ingredient needed is the Goldstino SUSY covariant derivatives which can be analogously defined as 
\bea
{\cal D}_\mu \lambda_A^\alpha &=& (A^{-1})_\mu~^\nu \partial_\nu \lambda^\alpha_A \cr
{\cal D}_\mu \bar\lambda_{\dot\alpha}^A &=& (A^{-1})_\mu~^\nu \partial_\nu \bar\lambda_{\dot\alpha}^A~,
\eea
so that their SUSY transformation is also that of the standard realization
\bea
\delta^Q(\xi,\bar{\xi})({\cal{D}}_\mu\lambda^\alpha_A)&=& \Lambda^\rho \partial_\rho \left( 
{\cal{D}}_\mu\lambda^\alpha_A \right) \cr
\delta^Q(\xi,\bar{\xi})({\cal{D}}_\mu\bar{\lambda}_{\dot\alpha}^A)&=& \Lambda^\rho \partial_\rho 
\left({\cal{D}}_\mu\bar{\lambda}_{\dot\alpha}^A\right).
\eea
Since they are singlets under any internal symmetry, all pure Goldstino terms are manifestly gauge invariant. On the other hand, recall that $\lambda_A$ and $\bar\lambda^A$ transform as fundamental representations of the $SU(N)_R$ symmetry while the Akulov-Volkov action is $R$-symmetric.

These standard realization building blocks consisting of 
the gauge singlet Goldstino SUSY covariant derivatives, 
${\cal D}_\mu \lambda_A,~ {\cal D}_\mu 
\bar\lambda^A$, the matter fields, 
$\phi^i$, their SUSY-gauge covariant derivatives, 
${\cal D}_\mu \phi^i$, and the field strength tensor, 
${\cal F}_{I\mu\nu}$, along with their higher covariant derivatives 
can be  combined to make SUSY and gauge invariant actions. These invariant 
action terms then dictate the couplings of the Goldstino which, in general,  
carries the residual consequences of the spontaneously broken extended supersymmetry.

A generic SUSY and gauge invariant action is thus constructed as 
\be
\label{IEFF}
\Gamma_{\rm eff}=\int d^4x \,\, \det A \,\, {\cal L}_{\rm eff}({\cal D}_\mu \lambda_A, 
{\cal D}_\mu \bar{\lambda}^A, \phi^i, {\cal D}_\mu \phi^i, {\cal F}_{I\mu\nu})
\ee
where ${\cal L_{\rm eff}}$ is any gauge invariant 
function of the standard realization basic building blocks. Using the 
nonlinear SUSY transformations 
$\delta^Q(\xi,\bar{\xi}) \, \det A = \partial_\rho (\Lambda^\rho \, \det A)$ and 
$\delta^Q(\xi,\bar{\xi}) {\cal L}_{\rm eff}= \Lambda^\rho \partial_\rho 
{\cal L}_{\rm eff}$, 
it follows that $\delta^Q(\xi,\bar{\xi}) \Gamma_{\rm eff}=0$. This structure is once again completely analogous to that found for the case of $N=1$ supersymmetry \cite{CL2},\cite{CL3}.

It proves convenient to expand $\cal{L}_{\rm eff}$ in this effective action in powers of the number of Goldstino fields which appear when covariant derivatives are replaced by ordinary derivatives and the Akulov-Volkov vierbein appearing in the standard realization field strengths are set to unity. The leading term in this expansion consists of all gauge and SUSY invariant operators 
made only from matter fields and their SUSY covariant derivatives.  Any Goldstino field which then appears arises only from higher dimension terms in the matter covariant derivatives and/or the field strength tensor. Denoting the non-Goldstino fields' Lagrangian by ${\cal L}_{\rm M}(\phi, D_\mu \phi, F_{\mu\nu})$,
then this leading term is given by the Lagrangian with the same functional form, but in which all derivatives replaced by SUSY 
covariant ones and the field strength tensor replaced by the 
standard realization field strength: ${\cal L}_{M}(\phi,{\cal D}_\mu \phi, {\cal F}_
{\mu\nu})$. Note that the coefficients of this term is fixed by the normalization of the 
gauge and matter fields, their masses and self-couplings; that is, the normalization of the Goldstino independent Lagrangian. The next term in this expansion of the effective Lagrangian begins with 
direct coupling of one Goldstino covariant derivative to the 
non-Goldstino fields.  The general form of these terms, retaining operators through mass dimension 6, is given by
$\frac{1}{f}[{\cal D}_\mu \lambda^\alpha_A (Q_M) _\alpha^{A~\mu} 
+ (\bar Q_M)_{A \dot\alpha}^{~\mu} {\cal D}_\mu \bar\lambda^{A\dot\alpha}]$, 
where $ (Q_M)_{ \alpha}^{A~\mu}$ and $(\bar Q_M) _{A\dot\alpha}^{~\mu}$ 
contain the pure non-Goldstino field contributions to 
the conserved gauge invariant supersymmetry currents with once again 
all field derivatives being replaced by SUSY covariant 
derivatives and the vector field strengths in the standard realization. 
That is, it is this term in the effective Lagrangian which, using 
the Noether construction, produces the Goldstino independent piece 
of the conserved supersymmetry current. 
This Lagrangian describes processes involving 
the emission or absorption of a single helicity $\pm \frac{1}{2}$ Goldstino. Finally the remaining terms in the effective Lagrangian all contain  
two or more Goldstino fields.  In 
particular, the next term in the exapnsion begins with the coupling of two Goldstino 
fields to matter or gauge fields. Retaining terms through mass dimension 8 
and focusing only on the $\lambda-\bar{\lambda}$ terms, we can write this term as  $\frac{1}{f^2}{\cal D}_\mu \lambda^\alpha_A {\cal 
D}_\nu 
\bar\lambda^{B\dot\alpha} (M_1)^{A~\mu\nu}_{B~\alpha\dot\alpha} +\frac{1}{f^2}
{\cal D}_\mu \lambda^\alpha_A \stackrel{\leftrightarrow}{\cal D}_\rho{\cal D}_\nu 
\bar\lambda^{B\dot\alpha} (M_2)^{A~\mu\nu\rho}_{B~\alpha\dot\alpha}  + \frac{1}{f^2}{\cal D}_\rho
\left[ {\cal D}_\mu \lambda^\alpha_A {\cal D}_\nu 
\bar\lambda^{B\dot\alpha}\right] (M_3)^{A~\mu\nu\rho}_{B~\alpha\dot\alpha}$, 
where the standard realization composite operators that contain matter 
and gauge fields are denoted by the 
$M_i$.  They can be enumerated by their operator dimension, Lorentz 
structure, field content and $R$-transformation behavior. Combining the various contributions gives
\bea
{\cal L}_{eff}&=&{\cal L}_{M}(\phi,{\cal D}_\mu \phi, {\cal F}_
{\mu\nu})\cr
&&+\frac{1}{f}[{\cal D}_\mu \lambda^\alpha_A (Q_M)_{\alpha}^{A~\mu} 
+ \bar (Q_M)_{ A\dot\alpha}^{~\mu} {\cal D}_\mu \bar\lambda^{A\dot\alpha}] \cr
&&+\frac{1}{f^2}{\cal D}_\mu \lambda^\alpha_A {\cal D}_\nu 
\bar\lambda^{B\dot\alpha} (M_1)^{A~\mu\nu}_{B~\alpha\dot\alpha} +\frac{1}{f^2}
{\cal D}_\mu \lambda^\alpha_A \stackrel{\leftrightarrow}{\cal D}_\rho{
\cal D}_\nu 
\bar\lambda^{B\dot\alpha} (M_2)^{A~\mu\nu\rho}_{B~\alpha\dot\alpha} \cr
&& + \frac{1}{f^2}{\cal D}_\rho
\left[ {\cal D}_\mu \lambda^\alpha_A {\cal D}_\nu 
\bar\lambda^{B\dot\alpha}\right] (M_3)^{A~\mu\nu\rho}_{B~\alpha\dot\alpha} +...  .
\eea

\noindent

This work was supported in part by the U.S. Department of Energy under grant DE-FG02-91ER40681 (Task B).


\end{document}